\documentclass[lettersize,journal]{IEEEtran}
\IEEEoverridecommandlockouts
\usepackage{cite}
\usepackage{amsmath,amssymb,amsfonts}
\usepackage{algorithmic}
\usepackage{algorithm}
\usepackage{array}
\usepackage{graphicx}
\usepackage{textcomp}
\usepackage{xcolor}
\usepackage[caption=false,font=normalsize,labelfont=sf,textfont=sf]{subfig}
\usepackage{stfloats}
\usepackage{url}
\usepackage{microtype}
\usepackage{cite}
\usepackage{todonotes}
\usepackage{tabularx}
\usepackage{xcolor}
\usepackage{hyperref}
\usepackage{tikzsymbols}
\usepackage{tikz}
\usepackage{pgfplots}
\PassOptionsToPackage{hyphens}{url}
\usetikzlibrary{shapes}
\usepackage{balance}

\usetikzlibrary{
    external,
    patterns,
    patterns.meta,
    arrows,
    positioning,
    chains,
}

\begin{document}

\title{Quantum Secure Anonymous Communication Networks}

\author{
    \IEEEauthorblockA{Mohammad Saidur Rahman},
    \IEEEauthorblockN{Stephen DiAdamo},
    \IEEEauthorblockN{Miralem Mehic}, and
    \IEEEauthorblockN{Charles Fleming}
    \thanks{Mohammad Saidur Rahman, Stephen DiAdamo are with the Cisco Quantum Lab in Los Angeles, USA and Munich, Germany. Mohammad Saidur Rahman is also with ESL Global Cybersecurity Institute, Rochester Institute of Technology, Rochester, USA. Miralem Mehic is with the University of Sarajevo, Sarajevo, Bosnia and Herzegovina. Charles Fleming is with Cisco Research, Los Angeles.}
    

}

\maketitle

\begin{abstract}

Anonymous communication networks (ACNs) enable Internet browsing in a way that prevents the accessed content from being traced back to the user. This allows a high level of privacy, protecting individuals from being tracked by advertisers or governments, for example. The Tor network, a prominent example of such a network, uses a layered encryption scheme to encapsulate data packets, using Tor nodes to obscure the routing process before the packets enter the public Internet. While Tor is capable of providing substantial privacy, its encryption relies on schemes, such as RSA and Diffie-Hellman for distributing symmetric keys, which are vulnerable to quantum computing attacks and are currently in the process of being phased out.

To overcome the threat, we propose a quantum-resistant alternative to RSA and Diffie-Hellman for distributing symmetric keys, namely, quantum key distribution (QKD). Standard QKD networks depend on trusted nodes to relay keys across long distances, however, reliance on trusted nodes in the quantum network does not meet the criteria necessary for establishing a Tor circuit in the ACN. We address this issue by developing a protocol and network architecture that integrates QKD without the need for trusted nodes, thus meeting the requirements of the Tor network and creating a quantum-secure anonymous communication network.

\end{abstract}


\section{Introduction}

In the modern digital era, there is a simultaneous unparalleled access to information and unprecedented threats to online privacy. The need for users to actively protect their privacy is increasingly important. In a recent example, the U.S. Congress repealed FCC rules that prohibited Internet Service Providers (ISPs) from trading users' browsing histories without  consent~\cite{nytimes2017}. Potential adversaries such as ISPs, autonomous systems, and wireless traffic sniffers can deduce our Internet activities and breach our privacy online. For instance, a recent FTC report found that many ISPs extensively collect and share user data, including sensitive information, with limited options for users to restrict this usage, raising significant privacy concerns~\cite{FTC2021}.

To maintain anonymity over the Internet, anonymous communication networks (ACNs) serve as tools against the dangers of third party Internet traffic monitoring. Beyond protecting individual users, ACNs play a pivotal role in global information dissemination. In countries where Internet censorship reigns, individuals often rely on tools like the Tor network~\cite{dingledine2004tor}, an ACN, to bypass restrictions to access information~\cite{khattak2014internet}. ACNs ensure that in an era where forces limit the flow of information, voices from the most suppressed regions can be heard. 

\begin{figure}[!t]
    \centering
    \includegraphics[scale=0.9]{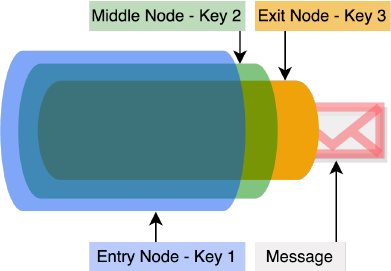}
    \caption{Three-layer encrypted message in a Tor network.}
    \label{fig:ontion-wrapping}
\end{figure}

ACNs differ from Virtual Private Networks (VPNs) in that they provide increased security and privacy by routing traffic through multiple nodes, making it difficult to trace online actions back to the user. On the other hand, VPNs provide an additional layer of security by encrypting traffic and typically routing it through a single node. However, since a VPN is based on a single node, there is a potential risk that the VPN node itself can be malicious.

One of the most popular ACNs used is the Tor network. The Tor network is renowned for enabling anonymous communication on the Internet and supports more than 8 million daily users worldwide. The network itself consists of over 8,000 relay nodes and around 2,000 bridge nodes distributed globally. Tor ensures user privacy through a process known as {\em onion routing}. When a user initiates a connection via Tor, their data does not travel directly to the destination. Instead, it is encrypted multiple times and sent through a randomly selected sequence of three nodes: an entry node, a middle node, and an exit node.

To establish this protected pathway, the Tor client first negotiates Advanced Encryption Standard (AES) session keys with each relay in a chosen path. This negotiation is guarded using Rivest–Shamir–Adleman (RSA) encryption, ensuring that these initial communications are secure. Additionally, the Diffie-Hellman (DH) key exchange is used, providing the exchange of cryptographic keys~\cite{dingledine2004tor}. Once the path and keys are established, the data packet is encrypted in layers corresponding to each relay's key. As the data passes through each relay---entry, middle, and exit---one encryption layer is peeled off. By the time the packet reaches its destination, it appears as though it originated from the exit node, obfuscating the true source (see  Fig.~\ref{fig:ontion-wrapping}).

While Tor's onion routing and cryptographic foundations have proven resilient against ``classical" threats, emerging quantum computational capabilities bring unique threats to protect against. Classical cryptographic primitives, like RSA and DH, relied upon by the Tor network, are grounded on the computational difficulty of specific mathematical problems. Yet, under the assumption of a quantum-powered adversary---a malicious third party with access to a quantum computer---these assurances weaken. A quantum computer able to perform Shor's algorithm could factorize large numbers in a relatively short amount of time~\cite{montanaro2016quantum}, thereby breaking the foundation in which the security presented by RSA encryption relies on. Similarly, the DH protocol, instrumental in ensuring secure key exchanges within Tor, would also face vulnerabilities in the face of a quantum adversary~\cite{dam2023survey}. 

In the face of emerging quantum technologies, the urgency to re-envision Tor's architecture for the post-quantum era has never been more pressing. With the cryptographic foundations that underpin Tor and many other digital systems potentially at risk, two pioneering defensive avenues emerge prominently: Post-Quantum Cryptography (PQC) and Quantum Key Distribution (QKD). PQC protocols, encompassing strategies like lattice-based, hash-based, and multivariate polynomial cryptography, are designed to withstand quantum attacks~\cite{bernstein2017post}. On the other hand, QKD is based on securing information via fundamental properties of quantum mechanics. QKD enables the distribution of a symmetric key between parties. The security of QKD comes from that eavesdropping attempts on the communication channel are detectable, serving as an alarm against adversaries~\cite{scarani2009security}. While PQC seeks to build algorithms tough enough to challenge quantum adversaries, QKD offers security grounded in the laws of physics, offering protection from any current or future computing platform.

Previous works have devised schemes that use PQC as the approach to distributing symmetric keys \cite{ghosh2015post, tujner2020qsor} for ACNs but a flaw of PQC protocols is that, so far, none of them have been able to prove what is known as ``everlasting security". Everlasing security is the idea that no matter the computation power of the adversary, the security of the key distribution protocol is guaranteed now and forever in the future. Indeed, QKD is not yet at a point where all information leaks are plugged, but new QKD protocols are appearing that are independent of the hardware that work towards this goal, one class of protocols being device-independent. 

We, therefore, investigate the use of QKD for designing a quantum-secure ACN based on the Tor design, ensuring that symmetric keys exchanged between a Tor client and the Tor nodes remain immune to classical and quantum threats. Specifically, our approach includes redesigning the key establishment and exchange mechanisms of Tor to incorporate QKD. Within our design, the keys exchanged between the Tor client and the Tor nodes achieve information-theoretic security. We further explain that using trusted nodes for key exchange, it is not enough to maintain anonymity between the Tor nodes with the same level of secrecy that Tor networks currently provide. Addressing these concerns, we propose a system design with the integration of QKD to maintain the properties of a Tor network and provide quantum-secure key exchange.

\section{Quantum Threats Against the Tor Network and the Realizability}
\label{sec:threat_model}

To break the encryption schemes used by the Tor network using quantum algorithms, the adversary needs access to a quantum computer. We define our primary attacker as a Quantum-Powered Adversary (QPA). Unlike conventional adversaries, those with access to classical computing technology, the QPA has the option to use advanced quantum computational resources, granting additional capabilities not available to classical computers. In this section, we start by stating the adversary's capabilities, the practicality of the threat, and follow with a review of the various possible threat vectors available to a QPA.

For the QPA, we assume they possess a quantum computer advanced enough to execute Shor's algorithm~\cite{montanaro2016quantum}. The use of Shor's algorithm allows quantum computers to factor large numbers efficiently, threatening RSA-based encryption schemes. Additionally, Grover's algorithm can search unsorted databases quadratically faster than classical counterparts~\cite{montanaro2016quantum}. Grover's algorithm poses a threat to symmetric key cryptography by effectively halving the security level, as it offers a theoretical quadratic speedup in brute force search. This means that a symmetric key of length $n$ bits would offer security equivalent to $n/2$ bits against an adversary with a quantum computer. 

To understand how Shor's algorithm works, we briefly review how a quantum computer functions. A qubit, the fundamental quantum computing resource, is the physical medium in which information is encoded into and manipulated, similar to a classical bit. A qubit has two major properties that give it additional abilities in comparison to a classical bit. The first is that a qubit has ability to be in a superposition of two states. Simply said, this means a qubit can be prepared in particular state-mixture, a particular measurement made to determine the qubit's state, and if repeated, with some probability, one can observe one qubit state and with complementary probability the other. The state is sometimes said to be in a probabilistic mixture and this allows for additional computational ability. The second key property is that multiple qubits can be {\em entangled} with each other. When qubits are entangled, the effect is that by measuring one (or more) of the entangled qubit system, it influences the measurement distribution of the remaining parts even though the other qubits are not modified explicitly.

\begin{figure}
    \centering    
    \includegraphics[scale=1]{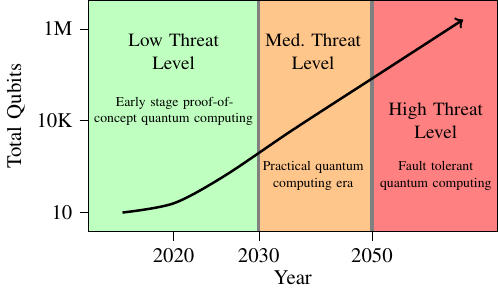}
    \caption{A predicted quantum threat level with time.}
    \label{fig:threat-vs-time}
\end{figure}

With superposition and entanglement, Shor's algorithm, in a simplified explanation, works in the following way. For a number $N$ to find the prime factors, the algorithm creates a large superposition state using (roughly) $2n$ qubits over all possibilities of $2n$-length bit strings (i.e., 00...0, 00...1, ..., 11...1), where $n$ is the smallest $n$ such that $2^n > N$. It performs a series of operations on the qubits that amplify the likelihood of measuring particular bit-strings from all of the $2n$-bit combinations, particularly those that are important for finding the factors of $N$. It simultaneously reduces the probability of the unimportant states being measured. Once the measurement statistics are gathered, post-processing on the most-likely states allows for finding the factors of $N$ with high probability.

To execute Shor's algorithm, the required quality of the quantum computer is estimated to be at a level that state-of-the-art quantum computers fall far short of. Current systems have a total number of qubits in the hundreds to thousands range, but to perform Shor's algorithm effectively, error-corrected quantum computers are needed which pose a significant engineering challenge. With that, the question of if quantum computers really are a threat arises. In \cite{OliverWyman2023} the authors estimate that the quantum threat is likely to be moderate after 2030 and the threat level will be high by 2050 when error-corrected quantum computers with millions of qubits exist, shown in Fig.~\ref{fig:threat-vs-time}. Given that the vast majority of our private communication relies on RSA, even if predictions are off, the U.S. government is of the opinion that the threat of quantum computers should be taken seriously and moving away from RSA-based cryptographic systems should be done. In particular, the Cybersecurity and Infrastructure Security Agency, the National Security Agency, the National Institute of Standards and Technology, and the Defense Advanced Research Projects Agency of the U.S. government have taken several initiatives and announced investment to protect systems from post-quantum threats~\cite{nsaquantum}. 

With the threat posed by quantum computers clarified, we can better understand the consequences of Tor's encryption scheme being compromised. The core strength of the Tor network lies in its layered approach to encryption, essential for maintaining the anonymity and security of its users. If a QPA successfully decrypts even one layer of this encryption scheme, it can trigger a cascading effect. The unraveling of even one layer, while others remain encrypted, can expose certain metadata or routing information, which can be exploited to decrypt subsequent layers. A QPA not only threatens individual privacy but also the broader use of Tor for secure communication in oppressive regimes. A breach by quantum methods could severely undermine free speech and information access. Thus, enhancing Tor's quantum resistance is crucial for protecting essential freedoms in the digital age.

\section{A Quantum-Resistant Tor Network}
\label{sec:QKDTor}

\begin{figure}[t]
\centering
\includegraphics[scale=1.1]{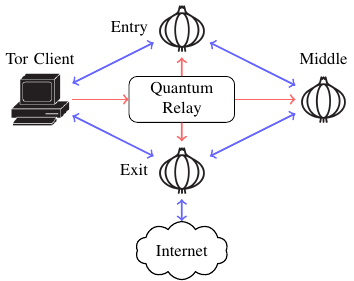}
\caption{QKD enabled and Quantum Secure Tor design. The Tor Client is classically connected (blue) to the Entry, Middle, and Exit node via a Tor circuit. The Tor Client connects to the input of the quantum relay with a quantum channel (red) and the output, depending on the frequency, is sent to the particular Tor node.}
\label{fig:QKDtor}
\end{figure}

In this section, we review the system requirements necessary to integrate QKD into a Tor network. We then describe our system design and protocol for implementing a quantum-secure ACN.

\subsection{System Requirements}

The system we propose must meet two key requirements: 1) It must adhere to the properties of the Tor network, and 2) It must employ a private key distribution protocol resistant to QPA attacks. The essential properties for a network to qualify as a Tor network are as follows: 1) Only the client should  be aware of the full path of the Tor circuit, which includes the entry, middle, and exit nodes; 2) Each node in the Tor circuit should only be aware of its immediate predecessor (sender) and successor (receiver), such that the entry node knows only the client and the middle node, the middle node knows only the entry and the exit nodes, and the exit node knows only the middle node and the final destination.

On the other hand, to generate symmetric keys using QKD, specialized quantum hardware both for the Tor client and the Tor nodes is required. The devices needed to perform QKD include a quantum source, quantum transmitters and receivers, quantum channels, and a source of randomness. To complete the QKD protocol after quantum transmission, the nodes need the ability to communicate classically for the post-processing steps. In a Tor network, as stated above, the middle and exit nodes cannot learn the identity of the Client, possibly revealed during quantum transmission. Thus, to generate quantum keys between the Client and the Tor nodes without directly connecting them, protecting the identity of the Client, a quantum relay device is needed. 

QKD is a symmetric key distribution protocol that can work in many ways---there are various protocols that achieve the same security, all using some form of quantum communication. The most commonly used QKD protocol found in consumer QKD hardware is BB84. BB84 works by preparing quantum states in a 0 or 1 state, or a superposition 0 or 1 state, randomly. It is not revealed to the receiver how the the state is prepared, and so the receiver can only guess the basis to measure the state. On error, they get a random output. Once the quantum transmission is complete, the sender and receiver must transmit classical information to finalize the QKD protocol. Before the protocol starts, they must firstly synchronize their preparation and measurement bases. Once the quantum transmission is complete, they perform error estimation and correction, and because error correction reveals information about the key, they must perform a privacy amplification step. The security of the protocol stems from the sender preparing the quantum state in one of two ways randomly. If there is an eavesdropper, they also can measure only in a random basis. They can wait for the post-processing steps to occur to try to extract key information. In this case, when estimating how much error occurred in transmission, the eavesdropper's attempts at extracting information will be detectable. The sender and receiver aborts key distribution in that case, throwing away any key material generated. 

After performing QKD, a symmetric key is generated between two hosts. An important aspect of any QKD network is how the keys are stored locally at the hosts. In our case, a non-trivial aspect arises when using QKD in an ACN setting, which is how the keys are stored and used. Generally, a key management system (KMS) stores keys in blocks of 128 bits, 256 bits, or similar. Each block has several metadata elements assigned to it, such as a unique ID, timestamp of generation, epsilon security, and others~\cite{mehic2022quantum}. 
However, in our case, we cannot associate the keys with the shared parties, as it would break the Tor Network requirements. For example, the exit node cannot know the client ID. Instead, we enforce that keys are given a network-wide unique identifier and that nodes store and transmit the IDs of the keys with their message. When a message arrives, the unique ID is accompanied by the message. If a node does not locally store that key, found by performing a lookup in the KMS, it cannot decrypt the message payload. This way, Tor nodes do not know who owns the other copy of the symmetric key.

\subsection{Network Design}\label{sec:network-design}

The primary objective of this design is to provide a theoretical guarantee of security against QPAs while meeting all system requirements. Here we outline the components of our proposed design as illustrated in Fig.~\ref{fig:QKDtor}. 

\subsubsection*{Tor Client}

At the start of communication, a Tor circuit must be constructed comprising of three nodes: Entry, Middle, and Exit. Only the Tor client can be aware of the entire path of the Tor circuit, and the Tor client's task is to generate a symmetric key with all three of the nodes. To generate the symmetric keys, two separate communication paths are required: one for classical communication and another for quantum communication. We describe how to establish these paths while masking the Client identity. 

\subsubsection*{Quantum Relay}

To establish a quantum connection between the Tor Client and the Tor Nodes, we use a quantum relay. A quantum relay is a device that can be designed as a passive device that has a single input of multiplexed signals. The signal, on arrival, can be routed to a particular output port passively. One such implementation was demonstrated in \cite{joshi2020trusted}, where the authors implement a dense wavelength division multiplexing to establish a trusted node–free eight-user metropolitan quantum communication network, with straightforward extension to very large and
complex local area quantum networks. 

The relay is therefore a device that does not need to process any classical information to make a routing decision and therefore none must be sent. The best the relay can do is determine the frequency of the signal, which can only reveal the signal destination. This information alone cannot be used to discover the Tor circuit, only the nodes that are acting as Tor nodes. Multiple Tor Clients transmit to various Tor circuits simultaneously, and knowing only the set of Tor nodes is not enough to determine the Tor Circuits. Another aspect of the quantum relay is that it does not measure the quantum information that arrives, keeping the information secret. If the relay were to attempt to measure the quantum information, it would be detectable by the QKD protocol and the key would not be used.

\subsubsection*{Entry, Middle, and Exit Nodes}

The entry, middle, and exit nodes each have two roles: first to establish symmetric keys with the client and second to continue the Tor protocol once the keys are established. A quantum communication channel's purpose is to transmit quantum states to establish secure keys, essential for safeguarding later transmitted data. On the other hand, classical communication is used for regular Internet activities.

Following Tor's design principles, every node in the Tor network is only aware of the node directly before and after it in the communication chain. This means the Tor client and the Entry node know each other's identities. The key created between the Tor client and the Entry node can use direct classical communication to establish a quantum channel without the involvement of other parties. In order to establish a quantum key, we place the quantum relay in the middle of network, forming a star-like subnetwork, although the network can be complex~\cite{joshi2020trusted}. As discussed previously, the quantum relay forwards quantum information without measuring it. The first step is for the Client and Entry to synchronize their QKD devices. The client requests a synchronization and the Entry responds with the frequency to transmit the signal through the switch so it is received, along with other common synchronization fields for QKD. Once synchronized, the client tunes its quantum source to the frequency and transmits to the relay to use the quantum channel and perform the QKD protocol. 

Setting up a secure key between the Tor Client and the Middle and Exit nodes is more challenging. The design needs to enforce that the Middle and Exit node remains unaware of the Tor Client's identity. In a QKD Network, it is often proposed to use trusted key-forwarding nodes. Here, forwarding the key through the Entry node, to the Middle node, for example, will not suffice. It would require the Entry node to have a copy of the key between the Client and Middle node, thereby allowing the Entry node to determine the Exit node's identity, breaking a Tor network principle. To address this, we again make use of the quantum relay. As already stated, the quantum signals containing key material do not contain classical information, and so once the quantum channel is established, the quantum information is sent anonymously. 

In a similar fashion, the client needs to synchronize its device with the Middle and Exit nodes while protecting its identity. By making use of the key is already generated with the Entry node, this can be done. We detail the protocol in the next subsection. Once the devices are synchronized, the Client can generate symmetric keys with the Middle and Exit nodes. From there the Tor protocol can run as it is defined with the assurance that all symmetric keys generated are secure.

\begin{figure*}
\centering
\includegraphics[scale=1.1]{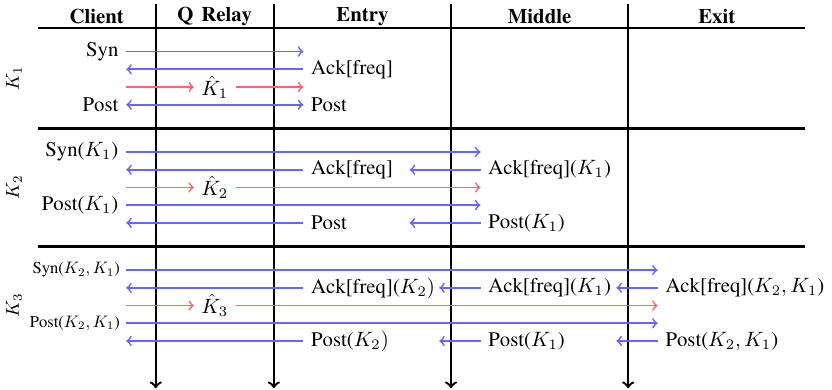}
\caption{Quantum secure key exchange in QKD enabled Tor network. The blue and red arrows represent classical and quantum communication respectively. The communicating parties are the Tor Client, the Quantum Relay, and the three Tor nodes. The classical messages are a Syn message (a synchronization request), Ack (an acknowledgment to the Syn) containing information of the QKD hardware for the Client, and Post are the post-processing messages for QKD. The quantum transmissions generate key material $\hat{K}_1, \hat{K}_2$, and $\hat{K}_3$.}
\label{fig:QKDtorKey}
\end{figure*}

\subsection{Key Exchange Protocol}\label{keyExchange}

A schematic representation depicting the key exchange protocol is shown in Fig.~\ref{fig:QKDtorKey}. The first step to any key distribution routine is authentication. There is open debate on how to perform authentication in a quantum safe manner, and we assume a combination of classical PQC and QKD can be used to secure the authentication step. The remainder of this section is the key generation protocol.

\subsubsection*{Key 1 -- Client and Entry}

In line with Tor's basic design, the Client and the Entry node can know each other's identity. Therefore, setting up a key for encrypted communication can be achieved using the QKD protocol as it is defined. This process does not require any additional parties as illustrated in Fig.~\ref{fig:QKDtorKey}. The client sends a synchronization request, Syn, getting the acknowledgment, Ack, from the receiver, containing its quantum transmission frequency. The client generates key material $\hat{K}_1$, transmitting through the quantum relay, then classically post-processed using a series of communication steps to complete the QKD protocol, generating $K_1$. 

\subsubsection*{Key 2 -- Client and Middle}

To establish a key between the Tor Client and the Middle node without exposing the Client, the communication steps work as follows. The Client must determine the transmission frequency and perform device synchronization with the Middle node, but it cannot reveal its identity. To mask its identity, it prepares its Syn message, but uses $K_1$ to lock its true identity in the message, ensuring the key ID is in clear text. As a source of the message, it spoofs its identity to be the Entry node. 

The response to the Syn message is sent from the Middle node to the Entry node. The Entry node, who has access to $K_1$, can unlock the true receiver, the Client, and transmit the Ack onward. This process can repeat until the devices are synchronized. Once the devices are synchronized, the Client can send key material $\hat{K}_2$ to the Middle node, transmitting on the Middle node's frequency through the quantum relay. Once the key material $\hat{K}_2$ is sent, the post-processing steps are done in the same way as the synchronization, locking the true identity with $K_1$, and resulting in a key $K_2$. Note that reuse of $K_1$ under a one-time-pad encryption assumption is not possible, so we assume $K_1$ is large enough to accommodate these transmissions.

\subsubsection*{Key 3 -- Client and Exit}

The steps for establishing Key 3 between the Tor Client and the Exit node uses a similar approach used for establishing Key 2. To begin, a Syn message must be sent from the Client to the Exit node, but both the Client and the Entry node must be hidden from the Exit node. The only node that can be revealed to the Exit node is the Middle node. Therefore, the client locks, in an onion-like fashion, its true identity in a message using $K_1$, leaving the key ID clear. Next it locks the Entry node in $K_2$ into the message. It sets the message sender as the Middle node and transmits the Syn to the Exit node. 

The Exit node responds to this message to the Middle node. The middle node, holding $K_2$ can unlock the part of the message that says to send the response to the Entry node. In this stage, the sender of the Ack, the Exit node, must be hidden from the Entry node. Therefore, the sender is set to the middle node, and the true sender is locked via $K_2$. The message is then sent to the Entry node who can unlock the Client identity and relay the Ack message. Once synchronized, the material $\hat{K}_3$ is sent through the switch at the frequency of the Exit node. Once complete, the QKD post-processing steps will occur using the same approach as the Syn-Ack steps. As a result, the Client and Exit share $K_3$.

\subsection{Security and Privacy}\label{keyExchange}

The robust security of the overall system hinges on the key distribution phase, which is fortified by a protocol that remains impervious even under the assumption of adversaries with computational power surpassing current limits. QKD, our chosen method for symmetric key distribution, guarantees that even adversaries with computing capabilities beyond quantum computers will never breach the system's encryption. We rely on a passive quantum relay device to establish a sophisticated local area quantum network. This device is solely used to relay quantum signals to the intended destination without any measurement or disturbance of the quantum states. It is important to note that no accompanying classical information is transmitted through the relay, rendering it entirely untrusted. 

In QKD protocols, as we have previously discussed, classical messaging occurs after the raw key distribution stage. The Tor circuit, crucial to our system, is established sequentially. This sequential establishment allows us to utilize the already established Tor circuit nodes to uphold the second rule of the Tor network. Using the established keys with a one-time pad, we employ a locking mechanism to send messages through the existing Tor nodes in the circuit, gradually building up the full circuit. It is worth reiterating that no classical information traverses the quantum relay, and the client ensures that the classical messages adhere to the key rules for a Tor-based ACN.

Being untrusted is important to maintain the requirements of an anonymous communication network. We assume the quantum network used for key distribution and the ACN are separate networks. The first requirement of the Tor network is that only the Tor client knows the complete Tor circuit. If a trusted relay in the quantum network were used, it would be safe against leaking the secret keys, but it would break the first rule of the ACN. The trusted node would be aware of which nodes the Tor client is distributing keys to and thus could determine the Tor circuit. In our design, the quantum relay knows the possible Tor nodes, where there would be many, but it does not know the Tor client. This maintains the ACN.

\section{Practical Implementation Challenges}
\label{sec:discuss}

In this section, we list some practical challenges of implementing our design. First, point-to-point quantum communication is strongly distance limited. With only a quantum relay, over fiber optical communication the maximum distance for performing QKD would be less than 1,000 km, depending on the QKD protocol and fiber technology of choice, and a BB84-based protocol would be limited to roughly 100 km in fiber~\cite{huttner2022long}. The way to overcome distance limitations in QKD networks is to use trusted repeater nodes, which we have already mentioned that ACNs with QKD cannot be used, since they can gain access to the secret keys and thus break the requirements of a Tor network. The alternative is to use quantum repeaters, which poses many challenges in the current state.

The next is in user scaling. If we require each Tor node to have a unique frequency, then the quantum relay needs to be able to accommodate. For many users, likely a single quantum relay will not suffice, and a chain of switches will be needed. Each added switch adds additional loss, and so there will be a trade-off between the number of supportable users and the network radius. 

Finally, current QKD systems do not generate keys with high rates -- the higher quantities in the megabit/second range. Our protocol may require many bits of key to lock information during the protocol for secure communication. Therefore, many keys will be required which could significantly increase the duration to establish the Tor circuit, especially if using one-time pad encryption, in a lower quantum bit error rate regime, and using Cascade for error correction~\cite{mehic2017analysis}. By using AES-256 instead, this could reduce the resource requirements, but at a cost of weakened security. To increase the key rates is an engineering problem, and there efforts to develop QKD systems to increase the key rate significantly~\cite{liu2022advances}.  

\section{Conclusion}\label{sec:conclude}

In summary, current RSA based encryption is under threat by future quantum computers. Our proposed QKD-based Tor network architecture provides information-theoretic security against such quantum-powered adversaries. Consequently, the adoption of QKD within Tor aims to create a robust defense mechanism, safeguarding the transmitted keys from interception or decryption by quantum technologies. Our approach involves a novel use of QKD and a quantum relay for establishing Tor circuits. Furthermore, our design incorporates an information locking protocol to ensure compatibility with Tor's existing fundamental protocol, offering a quantum-resistant solution that upholds the network's anonymity and privacy principles. While this design is applicable to metropolitan sized networks,
scaling up our QKD-integrated Tor network design is a direction for future research. This advancement represents a crucial step in future-proofing anonymous communication networks against the evolving landscape of quantum computing.

\bibliographystyle{IEEEtran}

\balance

\end{document}